\documentstyle{mn}

\newcommand{\Meszaros}{M\'esz\'aros}
\newcommand{\Paczynski}{Paczy\'nski}
\newcommand{\beq}{\begin{equation}}
\newcommand{\eeq}{\end{equation}}
\newcommand{\half}{\frac{1}{2}}
\newcommand{\third}{\frac{1}{3}}
\newcommand{\smaller}{\scriptscriptstyle}
\newcommand{\Gs}{\Gamma_{\rm s}}
\newcommand{\Egzk}{E_{\rm\smaller GZK}}
\newcommand{\Egrb}{{\cal E}_{\rm\smaller GRB}}
\newcommand{\Ei}{E_{\rm i}}
\newcommand{\Ef}{E_{\rm f}}
\newcommand{\Gr}{\Gamma_{\rm r}}
\newcommand{\br}{\beta_{\rm r}}
\newcommand{\mud}{\mu_{{\smaller\rightarrow}\rm d}}
\newcommand{\muu}{\mu_{{\smaller\rightarrow}\rm u}}
\newcommand{\bs}{\beta_{\rm s}}
\newcommand{\thd}{\theta_{{\smaller\rightarrow}\rm d}}
\newcommand{\thu}{\theta_{{\smaller\rightarrow}\rm u}}
\newcommand{\bbetai}{\bbeta_{\rm i}}
\newcommand{\bbetaf}{\bbeta_{\rm f}}
\newcommand{\Bprp}{B_{\smaller\perp}}
\newcommand{\Bpll}{B_{\smaller\parallel}}
\newcommand{\lc}{\ell_{\rm c}}
\newcommand{\rg}{r_{\rm g}}
\newcommand{\ran}{\rangle}
\newcommand{\lan}{\langle}
\newcommand{\zs}{z_{\rm s}}
\newcommand{\tu}{t_{\rm u}}
\newcommand{\td}{t_{\rm d}}
\newcommand{\wcp}{\omega_{{\rm c} \smaller\perp}}
\newcommand{\wc}{\omega_{\rm c}}
\newcommand{\vA}{v_{\rm A}}
\newcommand{\Rs}{R_{\rm s}}
\newcommand{\Rd}{R_{\rm d}}
\newcommand{\mpr}{m_{\rm p}}
\newcommand{\cE}{{\cal E}}
\newcommand{\eV}{\mbox{eV}}
\newcommand{\erg}{\mbox{erg}}
\newcommand{\cm}{\mbox{cm}}
\newcommand{\ecr}{e_{\rm\smaller CR}}
\newcommand{\yr}{\mbox{yr}}
\newcommand{\s}{\mbox{s}}
\newcommand{\gw}{\gamma_{\rm w}}
\newcommand{\Edot}{\dot{E}}
\newcommand{\Ngj}{\dot{N}_{\rm\smaller GJ}}
\newcommand{\mi}{m_{\rm i}}
\newcommand{\me}{m_{\rm e}}
\newcommand{\hi}{h_{\rm i}}
\newcommand{\hpm}{h_{\smaller \pm}}
\newcommand{\xii}{\xi_{\rm i}}
\newcommand{\Rb}{R_{\rm b}}
\newcommand{\dr}{{\rm d}}
\newcommand{\Emax}{E_{\rm max}}
\newcommand{\Emin}{E_{\rm min}}

\title[UHECR acceleration by relativistic blast waves]
      {Ultra-high-energy cosmic ray acceleration by
       relativistic blast waves}

\author[Y.A. Gallant and A. Achterberg]
       {Yves A. Gallant$^1$ and Abraham Achterberg$^{1,2}$ \\
        $^1$Astronomical Institute, Utrecht University,
            P.O.\ Box 80\,000, 3508 TA Utrecht, The Netherlands \\
        $^2$Center for High Energy Astrophysics, Kruislaan 403, 
           1098 SJ Amsterdam, The Netherlands}

\begin{document}

\maketitle

\begin{abstract}
   We consider the acceleration of charged particles at
the ultra-relativistic shocks, with Lorentz factors
$\Gs \gg 1$ relative to the upstream medium, arising in relativistic
fireball models of gamma-ray bursts (GRBs).  We show that
for Fermi-type shock acceleration, particles initially
isotropic in the upstream medium can gain a factor of order
$\Gs^2$ in energy in the first shock crossing cycle, but that
the energy gain factor for subsequent shock crossing cycles
is only of order 2, because for realistic deflection processes
particles do not have time to re-isotropise upstream before
recrossing the shock.

   We evaluate the maximum energy attainable and the efficiency
of this process, and show that for a GRB fireball expanding into
a typical interstellar medium, these exclude the production of
ultra-high-energy cosmic rays (UHECRs), with energies in the
range $10^{18.5}$--$10^{20.5}$\,eV, by the blast wave.  
We propose, however, that in the context of neutron star binaries
as the progenitors of GRBs, relativistic ions from
the pulsar wind bubbles produced by these systems could be
accelerated by the blast wave.  We show that if the known binary
pulsars are typical, the maximum energy, efficiency, and spectrum
in this case can account for the observed population of UHECRs.
\end{abstract}


\section{Introduction}
 
 The observation of some hundred cosmic ray events to date
 with energy above the
 Greisen--Zatsepin--Kuz'min cutoff,
 $\Egzk \simeq 10^{19.5}$\,eV,
 (e.g.\ Takeda et al.\ 1998, and references therein)
 has sparked renewed interest in their origin. 
 Ultra-high-energy cosmic rays (UHECRs), with energies in the range  
 $E \sim 10^{18.5}$--$10^{20.5}$\,eV,
 are generally believed to be extragalactic in origin, based on their
 harder spectrum, isotropic arrival directions on the sky and 
 the fact that they are not confined by the Galactic magnetic field.
One class of models involves {\em continuous} production of UHECRs
at large shocks, such as those associated with powerful radio galaxies,
or with ongoing large-scale structure formation in clusters of galaxies
(e.g.\ Norman, Melrose \& Achterberg 1995, and references therein).
A major difficulty with these models is that the number of possible
sources within the volume contributing to the flux above $\Egzk$,
which has a radius $D_{\rm max} \sim 50$--$100$ Mpc,
is too small to explain the observed number of independent events.
 
 An alternative model (Waxman 1995a; Vietri 1995; Milgrom \& Usov 1995)
 considers {\em impulsive} production in the sources of
 gamma-ray bursts (GRBs).  Observations of X-ray and optical afterglows
 of GRBs (van Paradijs et al.\ 1997; Metzger et al.\ 1997)
 have confirmed the cosmological origin of the phenomenon;
 the energy associated with each event is then thought to be
 $\Egrb \simeq 10^{51}$--$10^{53}$\,erg.
Waxman (1995a) and Vietri (1995) noted that assuming comparable
efficiencies for gamma-ray and UHECR production, the estimated GRB rate,
$Q_{\rm GRB} \sim 10^{-8}$ Mpc$^{-3}$ yr$^{-1}$, implies a flux
of UHECRs reaching Earth from within $D_{\rm max}$ remarkably
similar to the one observed.
Another, perhaps more compelling, argument is that
the dispersion in UHECR arrival times due to small-angle deflections
in the intergalactic magnetic field
implies that at any one time, enough GRB sources contribute to the
UHECR flux to account for the observed number of independent arrival
directions
(Miralda-Escud\'e \& Waxman 1996; Achterberg et al., in preparation).
 
 Relativistic fireball models for GRBs
 involve an ultra-relativistic blast wave with
 Lorentz factor $\Gs \simeq 10^{2}$--$10^{3}$ bounding
 the fireball (Rees \& \Meszaros\ 1992), 
 and internal mildly relativistic
 shocks ($\Gs \simeq 2$--$10$) due to unsteady outflows
 (Rees \& M\'esz\'aros 1994).
The quenching of interstellar scintillation observed in GRB radio
afterglows (Frail et al.\ 1997) confirms the relativistic expansion
of the source.
 Vietri (1995) proposed that these relativistic shocks
 can be sites of
 efficient particle acceleration up to energies exceeding $10^{20}$\,eV. 
 In particular, he argued that an ultra-relativistic shock with 
 Lorentz factor $\Gs$ will lead to an energy gain per crossing 
 cycle $\Ef/\Ei \simeq \Gs^2$, where $\Ei$ and $\Ef$ are
 the initial and final particle energies.
If such an energy gain could be obtained repeatedly at the
ultra-relativistic blast wave, UHECR energies would be reached
in only a few cycles.
 
 In this letter, we first consider the Fermi acceleration process
 at ultra-relativistic shocks in some detail; apart from the recent
 simulations by Bednarz and Ostrowski (1998), earlier calculations
 of such relativistic shock acceleration have concentrated on mildly
 relativistic shocks ($\Gs \la 10$).
 We distinguish between
 the energy gain in the initial and subsequent shock crossing cycles,
 and show that for physically realistic particle deflection processes
 upstream, the latter yield only an energy gain $\Ef/\Ei \sim 2$.
 We estimate the time scale for the acceleration process, and
 examine the maximum energy attainable, as well as the global
 energetics, for UHECR production scenarios at relativistic blast
 waves, first in a typical interstellar medium, and then in the
 plausible environment of a pulsar wind bubble.

\section{Ultra-Relativistic Shock Acceleration}

   We consider an ultra-relativistic shock of Lorentz
factor $\Gs \gg 1$ relative to the upstream medium.
Assuming the fluid is weakly magnetised, the shock jump
conditions then imply that the shock velocity relative to the
downstream medium reduces to $c/3$, while the relative Lorentz
factor of the downstream and upstream media satisfies
$\Gr = \Gs / \sqrt{2}$ (e.g.\ Blandford \& McKee 1976).

\subsection{Energy gain: initial vs.\ repeated crossings}

   At non-relativistic shocks, the standard scenario of particle
acceleration (Krymskii 1977; Bell 1978; Axford, Leer \& Skadron 1978;
Blandford \& Ostriker 1978) assumes that both the upstream and downstream
media contain magnetic fluctuations which tend to isotropise particles
by elastic scattering in the respective fluid rest frames.

 Assuming that the same general principle operates at a relativistic
 shock, the energy gain per shock crossing cycle
 follows from relativistic kinematics (e.g.\ Peacock 1981; Achterberg
 1993). 
 For ultra-relativistic particles (of Lorentz factor $\gamma \gg 1$), it can
 be written in terms of initial and final energies $\Ei$ and $\Ef$ as:
 \begin{equation}
 \label{kingain}
 \frac{\Ef}{\Ei} = \Gr^{2} \: \left( 1 - \br \mud \right) \:
 \left(1 + \br \muu' \right) = \frac{\Ef'}{\Ei'} \; ,      
 \end{equation}  
where the first and second equalities apply to
shock crossing cycles which begin and end in the upstream and
downstream media, respectively.
 Here $\br$ is the velocity of the downstream medium (in units of the
 speed of light) relative to upstream, 
$\Gr$ is the associated Lorentz factor, and 
the quantities $\mud$ and $\muu'$ measure the cosine of the angle
between the particle velocity and the shock normal, when the particle
crosses the shock into the downstream and upstream respectively.
 We use the convention that the shock normal points into the
 upstream medium, so that $\muu' > 0$.
Primed and unprimed quantities are respectively measured in the
downstream and upstream rest frames. 
The only assumption is that scattering is elastic in the local
fluid frame.

   Kinematics require that $1 \ge \muu' > \bs' = \third$, so that
the factor $(1 + \br \muu')$ in (\ref{kingain}) is of order unity.
If $\mud$ is more or less isotropically distributed, as would be the
case for a population of relativistic particles already present in the
undisturbed upstream medium, the factor $(1 - \br \mud )$
is also of order unity, and energy gains $\Ef/\Ei \sim \Gs^2$
can be achieved, as envisioned by Vietri (1995).
A similar conclusion holds for particles
which are initially non-relativistic upstream.

   For all but the initial shock crossing into the downstream medium,
however, the distribution in $\mud$ for a relativistic shock will be highly
anisotropic (Peacock 1981).
For an ultra-relativistic particle with Lorentz factor $\gamma \gg \Gs$,
the Lorentz transformation of the incidence angle reduces to
$\mu' \approx (2 - \Gs^2 \theta^2) / (2 + \Gs^2 \theta^2)$
when $\theta \equiv \cos^{-1} \mu \ll 1$.  The kinematic condition
$\muu' > \third$ is thus equivalent to $\thu < 1/\Gs$, which defines
the shock `loss cone' in the upstream frame.
We show below that for physically realistic deflection
processes upstream, the particle cannot be deflected very far beyond
this loss cone before the shock overtakes it, so that the angle
$\thd \sim 1/\Gs$ as well.  This severely limits the energy gain
attainable for all but the initial shock crossing:
when $\thd \ll 1$, the downstream energy gain in (\ref{kingain})
reduces to
\beq
\label{ratio}
       \frac{\Ef'}{\Ei'} =
       \frac{1 + \br \muu'}{1 + \br \mud'} \approx
       \frac{2 + \Gs^{2} \thd^{2}}
       {2 + \Gs^{2} \thu^{2}} \; ,
\eeq
which is of order unity if $\thd \sim 1/\Gs$.

\subsection{Upstream dynamics}

We consider two mechanisms for the upstream deflection of particles
needed to allow repeated shock crossings: regular deflection by a
large-scale magnetic field, and scattering by small-scale magnetic
fluctuations.

\subsubsection{Regular deflection}
 
   We first examine the case where the magnetic field may be
considered uniform over the region sampled by the particle in
its excursion upstream.  We take the shock velocity to define
the $z$-direction, and without loss of generality assume that
the magnetic field lies in the $x-z$ plane, 
$\bmath{B} = (\Bprp \, , \, 0 \, , \, \Bpll)$,
with $q \Bprp \ge 0$.
 
The equation of motion for the velocity $\bbeta = \bmath{v} / c$
of an ultra-relativistic particle of charge $q$ 
can be solved approximately by expanding $\beta_z = \mu \approx
1 - \half \theta^2$ to lowest
order in $\beta_x$ and $\beta_y$, as $\theta^2 \approx \beta_x^2
+ \beta_y^2 = {\cal O}(1/\Gs^2)$ throughout the particle's
orbit upstream.  Then as long as $\Bprp \gg \Bpll / \Gs$,
which will be the case for all but a fraction $\sim 1/\Gs^2$ of
possible magnetic field orientations, we can ignore the terms in
$\Bpll$ in the equation of motion.  If we denote by $\bbetai$
and $\bbetaf$ the ingress and egress particle velocities, we find
that $\beta_{{\rm f}x} = \beta_{{\rm i}x}$ and
\begin{equation}
\label{final}
        \beta_{{\rm f}y} = - \half \beta_{{\rm i}y} +
        \sqrt{\displaystyle \frac{3}{\Gs^{2}} - 3 \beta_{{\rm i}x}^{2}
        - \frac{3}{4} \beta_{{\rm i}y}^2} \; .
\end{equation}
Since $\beta_{{\rm i}x}^2 + \beta_{{\rm i}y}^2 < 1/\Gs^2$, eq.\
(\ref{final}) implies
\beq
        1 < \Gs \thd \leq 2
        \qquad \Longleftrightarrow \qquad
        \third > \mud' \geq -\third \; .
\eeq

   The energy gain ratio can then be obtained by substituting $\thd^2
\approx \beta_{{\rm f}x}^2 + \beta_{{\rm f}y}^2$ in (\ref{ratio}).
It has its maximum at $\beta_{{\rm i}x} =0$ and 
$\Gs \beta_{{\rm i}y} \approx -0.27$, where $\Ef'/\Ei' \approx 2.62\,$.
 
\subsubsection{Direction-angle scattering}

   We now consider the case where the magnetic field upstream is
not uniform but irregular.  In the simplest such model, the field
consists of randomly oriented magnetic
cells with field amplitude $B$ and radius (correlation length) $\lc$.
If the particle's gyration radius, $\rg \equiv E / q B$, satisfies
$\rg/\Gs \le \lc$, it will be deflected sufficiently to
recross the shock within a single cell, and the regular deflection
regime above will apply.  In the opposite case, $\rg/\Gs \gg \lc$,
the particle's momentum direction will diffuse in time, with angular
diffusion coefficient $D_0 = c \, \lc / (3 \rg^2)$ (Achterberg
et al., in preparation).

   In this scattering regime, a particle initially crossing the shock
at an angle $\thu$ will, after a time $t$, be at an average distance
upstream relative to the shock given by
\beq
\label{diffdist}
        \lan z(t) \ran - \zs(t) \approx c\,t \left( \frac{1}{2 \Gs^2}
        - \frac{\thu^2}{2} - D_0 t \right),
\eeq
where we have used the fact that $\theta = {\cal O}(1/\Gs)$ and
expanded to lowest order in $1/\Gs$.  Defining the typical upstream
residence time $\tu$ as the solution of $\lan z(\tu) \ran = \zs(\tu)$,
the typical downstream recrossing angle may be estimated as
\beq
\label{diffangle}
        \lan \thd^2 \ran \simeq  \lan \theta^2(\tu) \ran
                         \approx \frac{2}{\Gs^2} - \thu^2.
\eeq

   The typical direction angle upon recrossing the shock,
$\lan \thd^2 \ran^{1/2}$,
is thus again of order $1/\Gs$.  Substituting (\ref{diffangle})
into the energy gain formula (\ref{ratio}) shows that the energy
gain typically reaches its largest values when $\thu = 0$, where
$\Ef'/\Ei' \simeq 2$.  These results are approximate, rather than
exact averages, in that they neglect the correlation between
individual shock recrossing times and crossing angles.

\subsection{Acceleration time scale}

   As the energy gain per shock crossing cycle,
$\Delta E \equiv \Ef - \Ei$, is comparable to the initial energy 
$\Ei$, the acceleration time scale
will be of order the cycle time, which is the sum of
the typical upstream and downstream {\em residence times} $\tu$
and $\td$.  
   In the case of deflection by a uniform magnetic field, the
upstream residence time is the time required for the particle to
be deflected by an angle of order $1/\Gs$ [eq.\ (\ref{final})],
i.e.\ $\tu \sim (\Gs \wcp)^{-1} \equiv E / (q \Gs \Bprp c)$.

   The downstream residence time will depend on the downstream
scattering process, but if we assume Bohm diffusion, it can be
roughly estimated as the gyrotime, i.e.\ $\td' \sim \wc'^{-1}
\equiv E' / (q B' c)$.  If the downstream magnetic field 
is simply due to the compression resulting from the shock jump
conditions, $B' \simeq \Bprp' = \sqrt{8} \Gs \Bprp$, then
taking into account the Lorentz transformation of the energy
for a typical particle emerging upstream ($E' \sim E / \Gs$)
and the transformation of the time interval between two events
occurring at the shock ($\td' \simeq \td / \Gs$), one finds that
$\td \sim \tu$.
   Moreover, turbulence downstream could amplify the magnetic
field well above the value resulting from simple compression;
assuming it reaches equipartition with the thermal pressure
downstream, one can show that in this case $B' \sim (c / \vA)
\Gs B$, where $\vA$ is the upstream Alfv\'en speed.  This
yields a correspondingly shorter downstream residence time:
\beq
\label{tdown}
\td \sim \frac{E}{q \, c^2 \Gs \sqrt{4\pi \rho}},
\eeq
where $\rho$ is the upstream mass density.

   The full shock crossing cycle time is thus dominated by $\tu$ in
all cases of interest.  In the case of scattering upstream, where
$\rg \gg \lc$, solving (\ref{diffdist}) yields a typical upstream
residence time $\tu \sim \rg^2 / ( c \Gs^2 \lc )$.  This may be
combined with the regular deflection case in the single expression
\beq
\label{tup}
        \tu \sim \frac{E}{q \Gs B c} \times {\rm max} \left( \: 1 \; ,
                 \frac{\rg}{\Gs \lc} \right) .
\eeq
For a given upstream magnetic field amplitude $B$, the regular
deflection regime thus constitutes an absolute lower limit on the
acceleration time for repeated shock crossings.

\section{Fireballs in the general ISM}

   We now review the properties of the relativistic blast wave
driven into the surrounding interstellar medium in fireball
models of gamma-ray bursts.  After an acceleration phase, an initially
radiation-dominated fireball enters a relativistic `free expansion' phase,
driving a blast wave with the approximately constant Lorentz factor
$\Gs \simeq \sqrt{2} \eta$, where $\eta \equiv \Egrb / (M c^2)$,
$M$ being the baryonic
mass in which the fireball energy $\Egrb$ is initially deposited
(Piran, Shemi \& Narayan 1993; \Meszaros, Laguna \& Rees 1993).
The parameter $\eta$ is generally assumed to lie in the range
$10^2$--$10^3$.

If cooling of the shocked material can be ignored, this is followed
by an adiabatic deceleration stage
in which the blast wave Lorentz factor decreases with
radius $\Rs$ as $\Gs \propto \Rs^{-3/2}$ (Blandford \& McKee 1976).
The transition between the free expansion and deceleration phases
occurs roughly at the radius $\Rd$ at which the energy in the swept-up
surrounding material becomes comparable to $\Egrb$:
\beq
\label{decradius}
        \Rd \simeq \left( \frac{3}{4\pi} \frac{\Egrb}{\eta^2 e}
                  \right)^{1/3} ,
\eeq
where $e$ is the total energy density of the surrounding material;
for a typical interstellar medium of number density $n$, this is
simply the rest-mass energy $e \approx n \, \mpr c^2$.
The subsequent evolution of $\Gs$ is then given by
\beq
\label{decgamma}
        \Gs(\Rs) \simeq \sqrt{2}\,\eta \left(\frac{\Rs}{\Rd}\right)^{-3/2}
        \simeq \left( \frac{3}{2\pi} \frac{\Egrb}{e} \right)^{1/2} \:
        \Rs^{-3/2} \: ,
\eeq
and thus once $\Rs > \Rd$ the behaviour of $\Gs$ is independent of
the value of $\eta$.

   We now consider various scenarios for the acceleration of UHECRs
by the relativistic blast wave of a fireball expanding into the general
interstellar medium.

\subsection{Maximum energy for Fermi acceleration}

   In order for particles to be accelerated beyond the initial boost
by shock acceleration of the Fermi type, repeated shock crossings must
occur.  The acceleration time, which we argued above is dominated by
the upstream residence time $\tu$, must therefore be shorter than the
age of the blast wave measured in the upstream frame, $\Rs / c$.
Using (\ref{tup}), the maximum energy for which this can be the case
is
\beq
\label{limit}
        E \la q B \Gs \times \mbox{\rm min} \left( \Rs ,
                              \sqrt{\lc \Rs} \right) .
\eeq
The most favourable regime energetically is that of regular deflection
upstream, which requires $\lc \ga \Rs$; in what follows we will assume
that this is the case.
Note that the maximum energy (\ref{limit}) is larger by a factor $\Gs$
than that
resulting from a simple geometrical comparison of the gyration radius
with $\Rs$.  This is due to the fact that a particle typically only
executes a fraction $\sim\! \Gs^{-1}$ of a Larmor orbit upstream
before recrossing the shock.

   The behaviour of $\Gs$ as a function of $\Rs$
implies that the highest energy in (\ref{limit}) is reached at the
transition between the free expansion and deceleration phases, $\Rs
\simeq \Rd$.  Using the definition (\ref{decradius}) of $\Rd$, we
obtain:
\beq
        E \la 5 \times 10^{15} \: Z \, B_{-6} \cE_{52}^{1/3}
                                  \eta_3^{1/3} n_0^{-1/3} \: \eV \; ,
\eeq
for ions of charge $q = Z e$, where $B_{-6} \equiv B / \left(10^{-6}{\rm G}
\right)$, $\cE_{52} \equiv \Egrb / \left(10^{52} \,\erg\right)$, $\eta_{3}
\equiv \eta / 10^3$, and $n_0 \equiv n / \left(1 \,\cm^{-3}\right)$.
Given typical interstellar magnetic fields of a few microgauss and the
weak dependence on the other parameters,
this rules out the production of UHECRs by Fermi
acceleration at the unmodified external blast waves of relativistic
fireballs.

\subsection{Initial boost of galactic cosmic rays}

\subsubsection{Maximum energy}

  One remaining possibility for blast wave acceleration of UHECRs is that
they result from the initial shock crossing cycle energy boost
of a pre-existing upstream population of relativistic particles.
This is, in essence, the first of the two acceleration mechanisms
considered by Vietri (1995).

   This initial boost requires only the time $\td$ for the particle to be
scattered {\em downstream}, which if equipartition fields are generated can
be considerably shorter than $\tu$, yielding a correspondingly higher maximum
energy.  Requiring the downstream half-cycle time (\ref{tdown}) for a
particle of final energy $E$ to be shorter than the age of the fireball,
we obtain 
\beq
        E \la 7 \times 10^{20} \: Z \, \cE_{52}^{1/3} \eta_3^{1/3}
                                    n_0^{1/6} \: \eV \; .
\eeq
This process can thus attain UHECR energies, provided particles with
sufficient initial energy to be boosted in this range are present
upstream.  To reach $10^{20} \, \eV$, this requires relativistic
particles with energy above $\sim \! 10^{14} \, \eta_3^{-2} \, \eV$.

\subsubsection{Global energetics}

   The energy invested in boosting pre-existing relativistic particles
is of order $\Gs^2 \ecr$ per unit volume swept up by the blast wave,
where $\ecr$ is the upstream energy density of relativistic particles.
Meanwhile, the blast wave expends an energy $\sim \Gs^2 e$ per unit
volume in shock-heating the surrounding medium, which has upstream energy
density $e \approx n \mpr c^2$.  Thus the fraction of the fireball's
energy that can go into boosting cosmic rays is of order $f \sim \ecr/e$.
Taking the interstellar medium values in our Galaxy to be typical, $\ecr
\sim 1 \, \eV \, \cm^{-3}$ and $e \simeq 10^9 \, n_0 \, \eV \, \cm^{-3}$,
we see that this mechanism can only have a very low efficiency,
$f \la 10^{-9}$.

   The efficient yield of UHECRs required in the GRB hypothesis,
$f \ga 0.1$ (Waxman 1995a; Vietri 1995), could only be obtained if a large
fraction of the surrounding medium's energy density was in relativistic
particles.
This is in fact not implausible in the neutron star binary
merger scenario for GRBs, as we now argue.

\section{Fireballs in pulsar wind bubbles}

   The surrounding medium in which the relativistic fireball explodes
is probably one modified by the activity of the progenitor system prior 
to the burst event.
In the neutron star binary merger scenario for GRBs, these
progenitors are identified with the binary pulsar systems
observed in our own Galaxy (Narayan, Paczy\'nski \& Piran 1992).
The pulsar in these systems can be expected to emit a relativistic wind,
which over the lifetime of the binary can fill a large surrounding
volume with shocked relativistic plasma.  While this plasma will 
probably be
predominantly in the form of electron-positron pairs created in the
pulsar magnetosphere, it has been argued that pulsar winds must also
contain ions in order to account for the accelerated pair spectrum
observed in the Crab Nebula (Hoshino et al.\ 1992).  The Crab
Nebula's wisps provide additional observational evidence for this ion
component (Gallant \& Arons 1994).

\subsection{Maximum energy for boosted ions}

   We will assume the known binary millisecond pulsar systems
PSR~1913+16, PSR~1534+12, and PSR~2127+11C to be typical of GRB
progenitors.  The spiral-in times of these binaries due to
gravitational radiation are $\tau \sim 3 \times 10^7 \, \yr$
(Narayan et al.\ 1992), while the periods and period derivatives of
the millisecond pulsars observed in these systems yield spindown
luminosities $\sim \! 10^{33} \, \erg \: \s^{-1}$ and
characteristic times $\sim \! 10^8 \, \yr$ (Taylor, Manchester
\& Lyne 1993).  Thus these pulsars will continuously inject
relativistic particles into the surrounding medium over the lifetime
of the system, with an approximately constant wind luminosity.

   The relativistic flow of the pulsar wind will be thermalised in a
termination shock, and this shocked material will form a relativistic
plasma bubble in the interstellar medium.  At the ages involved,
we expect this pulsar wind bubble to be approximately isobaric
and in pressure equilibrium with the surrounding medium.
Once their directed kinetic energy is randomised by the shock, the
pairs may suffer energy losses by synchrotron radiation over time, but
these are negligible for the ions, which can only suffer adiabatic
expansion losses.  In an isobaric bubble these losses are negligible,
so that the ions approximately conserve their post-shock energy
throughout the bubble.

   The bulk Lorentz factor $\gw$ of the pulsar wind can be estimated
by assuming that the entire spindown luminosity $\Edot$
goes into a wind composed of electron-positron pairs and ions, with
number fluxes parameterised in terms of the
{\em Goldreich--Julian} flux $\Ngj$:
$\Edot = (\hi \mi + \hpm \me) \gw c^2 \Ngj$, where $\mi$ and $\me$ are
the ion and electron masses and $\hi$ and $\hpm$ are the ion and pair
{\em multiplicities}.
The wind energy per Goldreich--Julian charge is comparable to the
polar cap potential, and can be estimated as $\Edot / \Ngj \simeq
e (\Edot / c)^{1/2}$ (see, e.g., Michel 1991).
We expect the ions, although a minor component of the pulsar wind
particles by number ($\hi \ll \hpm$), to form the dominant component
by mass ($\hi \mi \ga \hpm \me$), so that the fraction of the wind energy
carried by them, $\xii \equiv [ 1 + \hpm \me / (\hi \mi) ]^{-1}$,
will be close to unity (Hoshino et al.\ 1992).  We will also
assume that the ions carry the magnetospheric return current,
so that $\hi \sim 1/Z$ (Gallant \& Arons 1994).  This then determines
the typical energy, $\mi \gw c^2$, of the thermalised post-shock ions.

   One can show that for an
initially isotropic distribution of relativistic particles upstream,
the average energy gain (\ref{kingain}) for boosted particles satisfies
$\frac{16}{9} \Gr^2 < \lan \Ef / \Ei \ran < \frac{8}{3} \Gr^2$,
the exact value depending on the typical angle $\lan \muu' \ran$
of those particles recrossing upstream.
Taking a factor $2 \Gr^2$ to be representative, the typical energy of
the ions boosted in the free expansion phase of the blast wave is
\beq
\Emax \simeq 2 \eta^2 \mi \gw c^2 \simeq 10^{20} \: \eta_3^2 \, Z \,
             \xii \, \Edot_{33}^{1/2} \: \eV \; ,
\eeq
where $\Edot_{33} \equiv \Edot / \left(10^{33} \,\erg \:\s^{-1} \right)$.
UHECR energies are thus naturally obtained is this scenario, for typical
values of the pulsar wind parameters, as long as $\eta \ga 10^3$.

\subsection{Spectrum and efficiency}

   Although the upstream ions have a relatively narrow energy
spectrum centered around $\mi \gw c^2$,
the boosted ions will have a broader distribution in energies
provided the blast wave decelerates within the pulsar wind bubble,
so that the boosting factor $\Gs^2$ is not constant.
The deceleration radius $\Rd$ is given by (\ref{decradius}), with $e$
the energy density of the relativistic plasma, while the pulsar wind
bubble radius $\Rb$ is obtained by equating the volume integral of $e$
with the total energy output of the pulsar.  It follows that
\beq
\frac{\Rd}{\Rb} \simeq 0.3 \: \cE_{52}^{1/3} \eta_3^{-2/3}
                            \Edot_{33}^{-1/3} \tau_7^{-1/3},
\eeq
where $\tau \equiv \tau_7 \times 10^7 \yr$ is the age of the system.
Thus for typical parameters, the relativistic fireball will decelerate
well within the pulsar wind bubble.

   Since $E \propto \Gs^2 \propto \Rs^{-3}$ in the deceleration phase,
and the number of ions swept up by the blast wave scales as
$\dr N \propto \Rs^2 \: \dr \Rs$,
the spectrum of the boosted ions will be
\beq
\label{spectrum}
\frac{\dr N}{\dr E} \propto E^{-2} \; .
\eeq
This is consistent with the observed UHECR spectrum (Waxman 1995b).
It can be shown that the same spectral index follows more generally
if the ion density is not uniform, as assumed here, but obeys a
power law in radius.

   The lower bound of the spectrum (\ref{spectrum}) follows from the
boosting factor $\Gs^2$ of the blast wave when it reaches the edge of
the pulsar wind bubble at $\Rb$, and is
\beq
\Emin \simeq 3 \times 10^{18} \: Z \, \xii \, \cE_{52}
               \Edot_{33}^{-1/2} \tau_7^{-1} \: \eV.
\eeq
Thus not only can a large fraction of the fireball energy go into
boosted ions, but these ions have a power-law spectrum extending
more or less exactly over the energy range where the extragalactic
UHECR component is observed.

\section{Summary}

   In this letter, we examined Fermi-type acceleration
at an ultra-relativistic shock, and found that particles with initially
isotropic momenta upstream can increase their energy by a factor of order
$\Gs^2$ in the initial shock crossing cycle.  In all subsequent shock
crossing cycles, however, we showed that the particle energy typically
only doubles.  This is due to the fact that particles do not have time
to re-isotropise upstream before being overtaken by the shock, which we
demonstrated in the specific cases of a large-scale ordered magnetic
field and of small-scale magnetic fluctuations.

   We argued that the maximum energy that can be reached by repeated
shock crossings at the unmodified relativistic blast wave from
a GRB fireball is well below the UHECR range.  Pre-existing
relativistic particles of sufficient energy can nonetheless be
boosted to UHECR energies in the first shock crossing cycle.
For a fireball expanding into a typical interstellar medium, where
galactic cosmic rays provide the seed particles, we showed, however,
that this process is too inefficient to account for UHECR production.

   We proposed that the blast wave instead expands into a pulsar wind
bubble produced by the progenitor system, a plausible hypothesis in
the neutron star binary merger scenario for GRBs.  We showed that for
parameters typical of the neutron star binary systems observed in our
Galaxy, relativistic ions in the pulsar wind bubble can be efficiently
boosted by the blast wave to energies exceeding $10^{20} \, \eV$.
We argued that these boosted ions would have an $E^{-2}$ spectrum,
extending down in energy to about $10^{18.5} \, \eV$.

\section*{Acknowledgements}

This work was supported by the Netherlands Foundation for Research in
Astronomy (ASTRON) project 781--71--050.

\end{document}